\documentclass[11pt]{article}
\usepackage{amsmath,amsfonts,amssymb,authblk,graphicx,braket,cite,multicol}   
\begin{document}

\noindent \textbf{\Large{Bi-photon spectral correlation measurements from a silicon nanowire in the quantum and classical regimes}}
\\
\\
\\
\small{
\noindent Iman Jizan$^{1,\ast}$, L. G. Helt$^{2}$, Chunle Xiong$^{1}$, Matthew J. Collins$^{1}$, Duk-Yong Choi$^{3}$, Chang Joon Chae$^{4}$, Marco Liscidini$^{5}$, M. J. Steel$^{2}$, Benjamin J. Eggleton$^{1}$ and Alex S. Clark$^{1}$
\\
\\
\noindent $^{1}$ Centre for Ultrahigh bandwidth Devices for Optical Systems (CUDOS), Institute of Photonics and Optical Science (IPOS), School of Physics, University of Sydney, New South Wales 2006, Australia\\
\noindent $^{2}$ CUDOS and MQ Photonics Research Centre, Department of Physics and Astronomy, Macquarie University, New South Wales 2109, Australia\\
\noindent $^{3}$ Laser Physics Centre, Australian National University, Canberra, Australian Capital Territory 2913, Australia\\
\noindent $^{4}$ NICTA-VRL, The University of Melbourne, Victoria 3010, Australia (now with Advanced Photonics Research Institute, GIST, Korea)}\\
\noindent $^{5}$ Dipartimento di Fisica, Universita degli Studi di Pavia, via Bassi 6, I-27100 Pavia, Italy\\
\noindent $^{*}$ Corresponding author email: imanj@physics.usyd.edu.au
\\
\\
\noindent The growing requirement for photon pairs with specific spectral correlations in quantum optics experiments has created a demand for fast, high resolution and accurate source characterization.  A promising tool for such characterization uses the classical stimulated process, in which an additional seed laser stimulates photon generation yielding much higher count rates, as recently demonstrated for a $\chi^{(2)}$ integrated source in A.~Eckstein \emph{et al.}, Laser Photon. Rev. \textbf{8}, L76 (2014). In this work we extend these results to $\chi^{(3)}$ sources, demonstrating spectral correlation measurements via stimulated four-wave mixing for the first time in a integrated optical waveguide, namely a silicon nanowire. We directly confirm the speed-up due to higher count rates and demonstrate that additional resolution can be gained when compared to traditional coincidence measurements. As pump pulse duration can influence the degree of spectral entanglement, all of our measurements are taken for two different pump pulse widths. This allows us to confirm that the classical stimulated process correctly captures the degree of spectral entanglement regardless of pump pulse duration, and cements its place as an essential characterization method for the development of future quantum integrated devices.

\begin{multicols}{2}
\section*{Introduction}

In the last decade the investigation of non-classical correlations between 
photons has been one of
the central topics in quantum optics. Quantum correlations between photon pairs
are a key resource for exceeding the technological limits
imposed by classical physics and play an integral part in many applications of
quantum optics including optical quantum computing~\cite{2007optical}, secure
communication over large distances~\cite{2009photonic,gisin2007quantum} and
quantum metrology~\cite{nagata2007}. The need for complex and precisely controlled 
correlated photon
states is the driving force behind the
development of new methods to accurately characterize correlated photon pair
sources.

\begin{figure*}[t]
\centering \includegraphics[width=11cm]{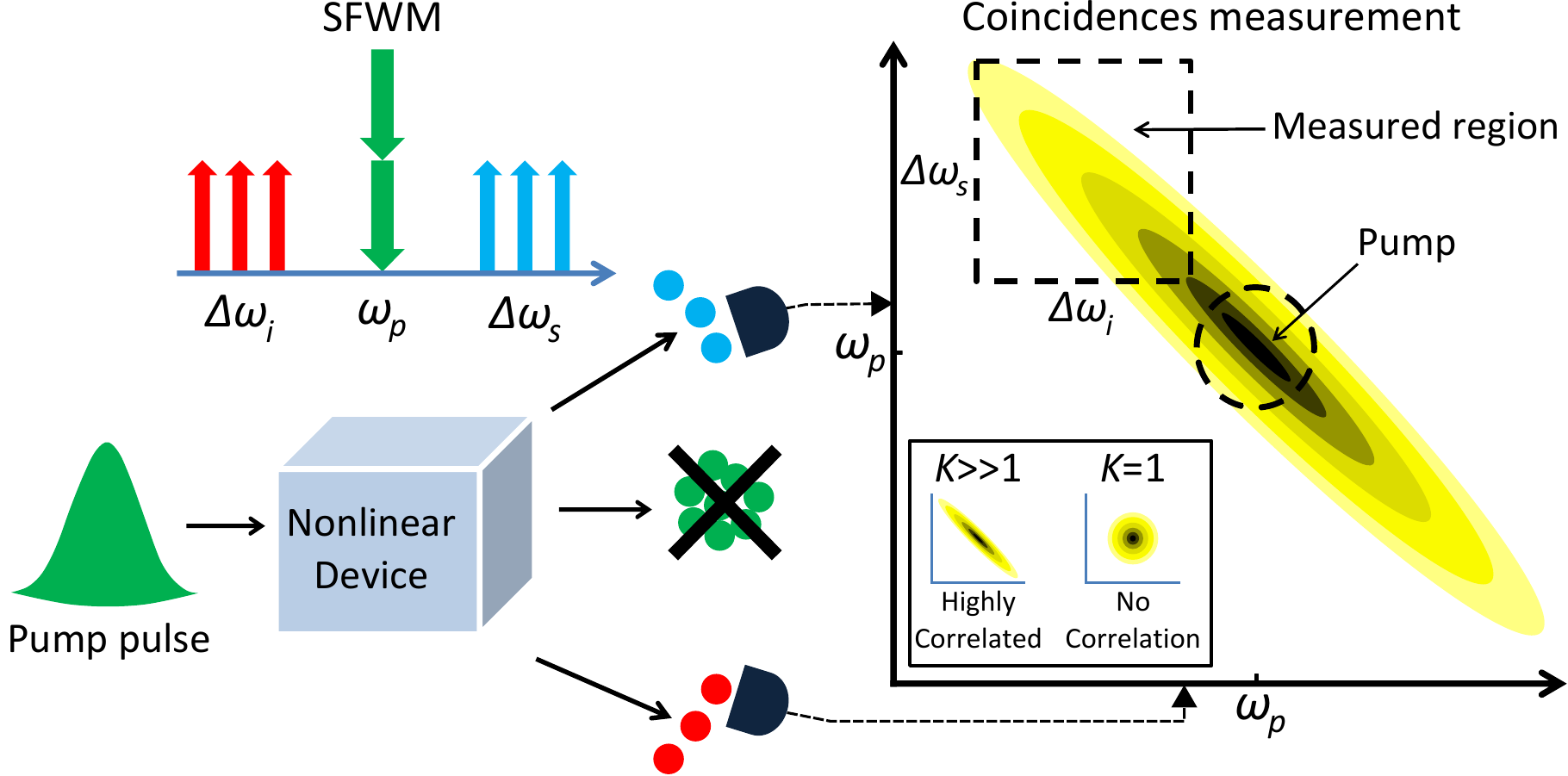} 
\caption{\footnotesize{
Schematic illustration of the SFWM process and a conventional JSI measurement. A
pulse is injected into a nonlinear device generating a signal and idler photon via the annihilation of two pump photons.The unconverted excess pump
photons are dropped. Measuring correlations across two detectors, the JSI is
obtained by recording the number of coincidences obtained at each specific idler
and signal frequency.}}
\label{Fig:Concept} 
\end{figure*}

Quantum correlations between photons can exist in many degrees of
freedom including polarization, time-bin, and energy. A common form of
correlation is energy-time correlation, also known as spectral entanglement,
which is particularly important in quantum communications
\cite{Nunn2013,Bernhard2013,Bessire2014,Lukens2014}. 
Spectrally entangled photons arise naturally~\cite{kwiat1995new} in
spontaneous parametric down-conversion (SPDC) and spontaneous four-wave mixing
(SFWM), in second order ($\chi^{(2)}$) and third order ($\chi^{(3)}$) nonlinear
materials respectively, as a result of the ultrafast nonlinear interaction and
energy-matching requirements. Until very recently, the most common method for
characterizing the degree of spectral entanglement of photon pairs has been
direct measurement of the joint spectral intensity (JSI). 
This function, defined formally below, is essentially the probability
distribution in frequency space for detecting pairs.
Generally, the JSI is obtained by
performing photon coincidence measurements in which the correlated
photon pairs are detected via a pair of single photon detectors over a range of
frequencies (see Fig.~\ref{Fig:Concept}). This has been performed a number of
times for SPDC using a tunable filter~\cite{mosley2008} or a highly dispersive
fiber~\cite{Gerrits11}, and for SFWM using either temporal dispersion in long
lengths of fiber~\cite{C2010}, monochromators~\cite{C2010,Spring13}, or
spatial mode separation of the signal and idler photons~\cite{jizan2014}. 
All of these schemes suffer from limitations in both achievable resolution and
acquisition times. The latter are typically large due to the need to measure
sufficient coincidences to obtain a satisfactory signal to noise ratio, usually
in the presence of low throughput. Clearly, rapid high-resolution characterization of the
spectral entanglement created in nonlinear pair sources is challenging.

\end{multicols}

\begin{multicols}{2}
To address this problem, Liscidini and Sipe~\cite{liscidini2013} introduced
a technique to reconstruct the JSI by performing stimulated nonlinear wave mixing.
This approach uses bright classical fields, exploiting the observation that, for a given pumping scheme and nonlinear device, the spontaneous and simulated frequency conversion response functions can be made mathematically identical. 
This was demonstrated recently in a $\chi^{(2)}$ device, namely an AlGaAs ridge
waveguide~\cite{Eckstein2014}, where the spontaneous process of SPDC was
compared to the stimulated process of difference frequency generation (DFG).
The experiment compared the JSIs obtained from SPDC via a temporal dispersion
method and stimulated DFG via an optical spectrum analyzer. The stimulated process using DFG produced a higher resolution in only a third of the
integration time. There has also been a single demonstration of reconstruction of the JSI via stimulated four-wave mixing (FWM) in a $\chi^{(3)}$ nonlinear device, namely a birefringent optical
fiber~\cite{Fang2014}. This work compared the JSI obtained from the
stimulated process to that taken by coincidence measurements on a similar, 
though not identical fiber, finding close resemblance.

Here we apply the stimulated process concept to an integrated $\chi^{(3)}$ nonlinear
device, in this case a silicon nanowire.
Silicon photonics is currently a leading platform for on-chip quantum
integrated circuits~\cite{Bonneau2012,Silverstone:2013}, due to the high
intrinsic $\chi^{(3)}$ nonlinearity, the possibility for dense integration,
mature fabrication methods, low losses and low cost~\cite{leuthold2010}. As
such, there is significant motivation to use integrated $\chi^{(3)}$ nonlinear
devices for generating quantum correlated photon pairs in the
telecommunications band \cite{Sharping2006,Harada2008,Clemmen2009,Xiong2011,Xiong2012,Engin2013,Clark2013Chad,Matsuda2013}
and to develop fast characterization techniques. In the following, we measure three JSIs, one via coincidence measurements from SFWM, and two via stimulated FWM using different detection methods. The classical stimulated FWM techniques produce fast and reliable results, which should be extensible to larger frequency ranges and directly applicable to many future, integrated nonlinear devices. Moreover, we  observe and compare the change in the spectral entanglement of photon pairs generated using two different pump pulse durations in the nonlinear device.

\section*{Formalism}

To understand the relationship between SFWM and stimulated FWM, we refer 
to Fig.\ref{Fig:Concept}, which illustrates the annihilation of two pump
photons resulting in the generation of a signal and idler photon of higher and
lower energy, respectively. 
For both processes, the frequencies must obey energy
conservation such that
\begin{equation}
2\omega_{p} = \omega_{s}+\omega_{i},
\end{equation}
where $\omega_{p}$, $\omega_{s}$ and $\omega_{i}$ are the pump, signal and
idler frequencies respectively. SFWM occurs in the absence of any
seed field, and instead relies on vacuum fluctuations to seed the
conversion of a pair of pump photons into correlated signal and idler photons.
In contrast, stimulated FWM involves a classical seed field in either the signal or
idler band and is much more efficient.  It forms the basis for parametric
oscillators~\cite{sharping2002optical,nakazawa1988modulational} and
ultra-broadband amplifiers~\cite{hansryd2002fiber}. 

For guided-mode co-polarized pair generation via SFWM, the two emitted photons occupy 
the same waveguide mode and 
the output state 
can be expressed as
\begin{equation}
\label{eq:ket}
\Ket{\psi} = \iint \text{d} \omega_{s}\text{d}\omega_{i} \,
F(\omega_{s},\omega_{i})\Ket{\omega_{s}}\Ket{\omega_{i}},
\end{equation}
where $\Ket{\omega}=\hat{a}^\dagger_{\omega} \Ket{\text{vac}}$ is 
the state containing a single photon in the waveguide mode at $\omega$.
Additionally,
\begin{equation}
\label{eq:jsatheory}
F(\omega_{s},\omega_{i})=\int \text{d}\omega \,
\alpha(\omega)\alpha(\omega_{s}+\omega_{i}-\omega)\phi(\omega_{s},\omega_{i},\omega) , 
\end{equation}
is known as the bi-photon wavefunction or 
joint spectral amplitude (JSA). Its squared modulus 
$|F(\omega_{s},\omega_{i})|^{2}$
defines the  JSI.
 The function $\alpha (\omega)$ is the complex
amplitude of the pump spectrum (with centre frequency $\omega_{p}$) and
$\phi(\omega_{s},\omega_{i},\omega)$ is the phase-matching function of the
waveguide which reflects the waveguide material and design properties. 
Knowledge of the JSA is thus equivalent to a complete description of
the quantum state according to \eqref{eq:ket}.

As a complex function of two variables, the JSA can be usefully analyzed in terms of 
the Schmidt decomposition, by which it is expressed  as a linear combination
\begin{equation}
F(\omega_{s},\omega_{i}) = \sum_{n}\sqrt{\lambda_{n}}f_{n}(\omega_{s})g_{n}(\omega_{i}),
\end{equation}
where $f_{n}(\omega_{s})$ and $g_{n}(\omega_{i})$ are each a complete set of
orthonormal functions, $\lambda_{n}$ are positive real numbers known as the
Schmidt magnitudes satisyfing $\sum_n \lambda_n=1$, and $n$ is an integer. This
can then be used to quantify the degree of entanglement in the system via the
Schmidt number $K=\left(\sum_{n}\lambda_{n}^2\right)^{-1}$ \cite{law2004analysis}.  For a completely
uncorrelated system the Schmidt magnitudes are $\lambda_{n=1}=1$ and
$\lambda_{n\neq1}=0$ so that $K=1$. However for a correlated system, 
multiple Schmidt magnitudes are nonzero so that $K>1$ (see
Fig.~\ref{Fig:Concept}: JSI plot inset). 

In fact, obtaining the full phase-dependent JSA for a bi-photon source is
experimentally challenging \cite{avenhaus2014time}, and experiments to date have focused on measuring
the JSI represented by $|F(\omega_s, \omega_i)|^2$, as we do here. However, this JSI measurement results in a loss of phase information when estimating $F(\omega_{i},\omega_{s})=\sqrt{|F(\omega_{i},\omega_{s})|^{2}}$. 
The Schmidt decomposition is not directly applicable to $|F(\omega_s, \omega_i)|^2$, and so
the Schmidt number $K$ is not strictly available from experiment.  However, a singular value decomposition (the matrix analog of the Schmidt decomposition,)
applied to the square root of the measured JSI,  $|F(\omega_s, \omega_i)|$,
does give a lower bound to the Schmidt number~\cite{Eckstein2014}, which
remains a useful characterisation of the source. In the following, we refer to the 
Schmidt number lower bound (SNLB),
with the symbol $\tilde{K}$.

\section*{Experimental Methods}
In this work we demonstrate three distinct
methods of obtaining JSIs from a $\chi^{(3)}$ nonlinear device using quantum,
singles-based and OSA measurements that provide progressive
improvements to the signal-to-noise ratio and measurement efficiency. We first
employ a high resolution spatial separation method~\cite{jizan2014} to determine the
JSI in the quantum regime by measuring the correlated photon pair coincidences
from SFWM. In the second experiment, we employ an additional narrow-band seed laser tuned across the signal band to stimulate classical FWM, and measure the spectrum of the generated idler field 
using a single photon detector. We refer to this as the singles-based approach.
Our final method again involves the measurement of the idler field generated via stimulated FWM, but in this case using a high resolution optical
spectrum analyzer (OSA). We refer to this as the OSA method. We compare our experimental methods for two different laser pump pulses and thus observe a change in the spectral entanglement of the photon
pairs generated in our nonlinear device.

\begin{figure*}[t!]
\centering
\includegraphics[width=15.5cm]{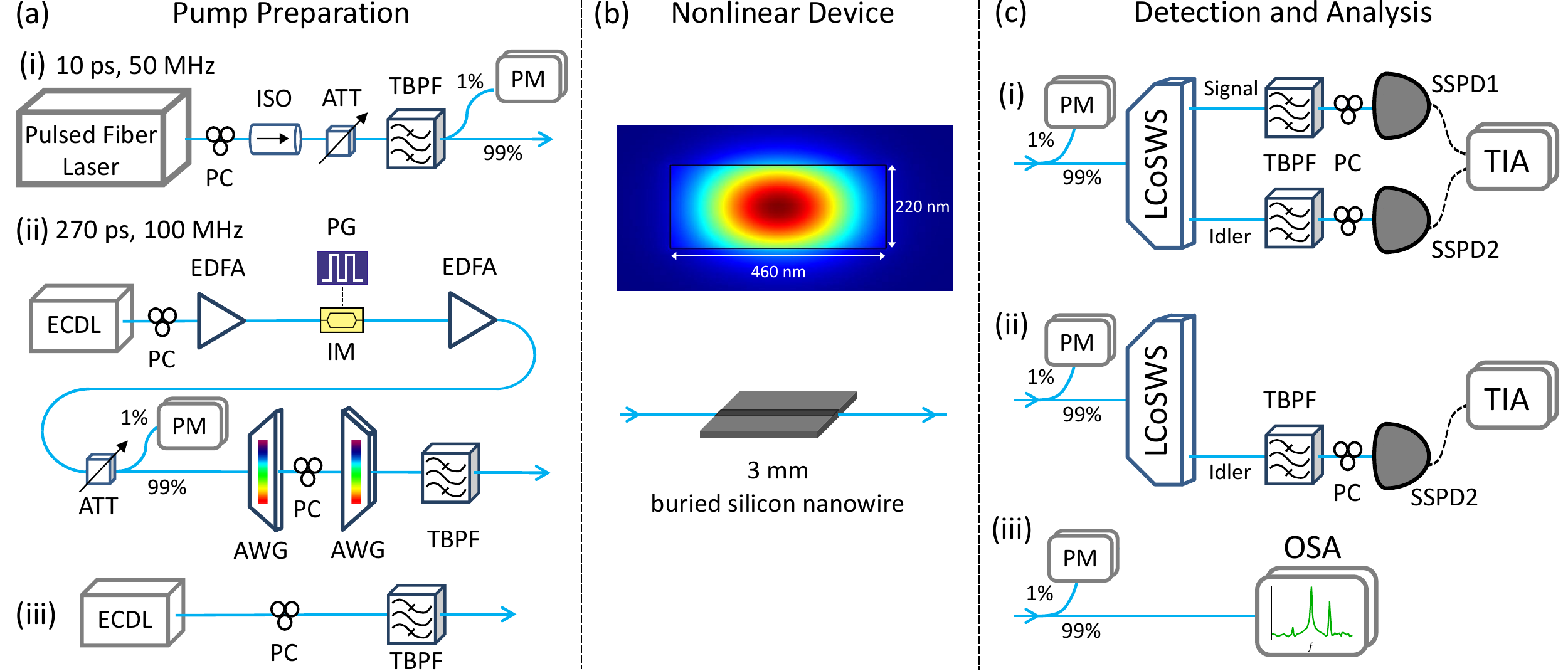}
\caption{\footnotesize{Schematic of all fiber-based JSI setups showing (a); the pulsed fiber laser, polarization controller (PC), optical isolator (ISO), variable attenuator, tunable band-pass filter (TBPF), power meter (PM), external cavity diode laser (ECDL), erbium-doped fiber amplifier (EDFA), pulse generator (PG), intensity modulator (IM), arrayed waveguide grating (AWG), (b); buried silicon nanowire (SiNW) and the simulated $|E|^2$ component of the fundamental TE mode, (c); liquid-crystal-on-silicon dynamically tunable filter (LCoSWS, Finisar WaveShaper), superconducting single photon detector (SSPD, Single Quantum - polarization sensitive), time interval analyser (TIA) and optical spectrum analyser (OSA).}
\label{Fig:ExperimentalSetup}}
\end{figure*}

The experimental setup is shown in Fig.~\ref{Fig:ExperimentalSetup}.  It
consists of three major parts required to perform SFWM coincidence
measurements and traditional stimulated FWM measurements in the nonlinear
device: the pump and seed laser preparation, the nonlinear device, and the
detection and analysis setup.
\end{multicols}

\begin{multicols}{2}
\subsection*{Pump and Seed Preparation} 
Figure~\ref{Fig:ExperimentalSetup}(a) shows the different pump and probe laser
preparations. The first pump source was a pulsed fiber laser (Pritel) centred
at 1550 nm (Fig.~\cite{Fig:ExperimentalSetup}(a)(i)) which produced 10~ps pulses
with a repetition rate of 50 MHz with 70.8 GHz spectral FWHM. The pulses passed
through a polarization controller (PC) to select TE polarization with respect
to the waveguide device, an isolator (ISO) to protect the laser and a variable
attenuator (ATT) to tune the input pump power. Residual cavity photons from the
laser were removed using a narrowband tunable band-pass filter (TBPF) before
entering a 99:1\% coupler to monitor the input power entering the nonlinear
device on a power meter (PM). 

The second pump source, shown in
Fig.~\ref{Fig:ExperimentalSetup}(a)(ii), used one channel of an external cavity
diode laser (ECDL) centred at 1550 nm which passed through a PC before being
pre-amplified by a low-noise erbium doped fiber amplifier (EDFA) to directly increase
the pump signal. The pump wave was modulated to 270 ps Gaussian
pulses at a repetition rate of 100 MHz by a lithium niobate intensity modulator
(IM, Sumitomo) driven by a pulse generator (PG, AVTech), resulting in a 10.4~GHz
spectral FWHM. The pump pulse stream was then amplified by a second EDFA and
subsequently filtered by two arrayed waveguide gratings (AWGs, JDSU) to remove
any amplified spontaneous emission noise. A PC was placed in between the two
AWGs to adjust the polarization such that the pump pulse was TE polarized in
the nonlinear device. 

Finally, the seed laser for the stimulated FWM experiments, shown
in Fig.~\ref{Fig:ExperimentalSetup}(a)(iii), used the second channel of the
ECDL which was also set to TE polarization using a PC. This channel of the ECDL was
computer controlled to repeatedly scan the higher-band channel over the desired
spectral detuning range from the pump, detailed below. 

\subsection*{Nonlinear Device}

As shown in Fig.~\ref{Fig:ExperimentalSetup}(b), our nonlinear device is a 3~mm
long silicon-on-insulator (SOI), 220 nm high by 460 nm wide buried silicon nanowire
(SiNW), providing an effective nonlinearity of approximately  $\gamma_\text{eff}
\sim 900~\mathrm{W^{-1}m^{-1}}$. To improve waveguide to fiber coupling efficiency, the TE-optimized waveguide was inverse tapered over a  200~$\mu\mbox{m}$ to a cross section of 220~nm high by 130~nm wide at the facet. The SiNW was photolithographcally fabricated and etched via reactive ion etching on an SOI
wafer with a 2~$\mu \mbox{m}$ upper-cladding silicon dioxide layer deposited via
plasma-enhanced chemical vapor deposition. The average power in front of the
waveguide was 4.9~mW and 790~$\mu\mathrm{W}$ for the 270~ps (100~MHz) and
10~ps (50~MHz) lasers respectively. These powers were set to generate the same
number of photon pairs per second in the device for the two laser pulse widths
and were below the threshold for two-photon absorption~\cite{Husko2013}.
Additionally, the average seed power in front of the waveguide was kept
constant at 36.5 $\mu$W. The TE propagation and coupling loss of the SiNW was approximately 2-2.5~dB/cm and 2-2.5~dB/facet respectively. The total loss between the input and output of the SiNW was 4.5~dB for all measurements.

\subsection*{Detection and Analysis}

Three different experimental setups were used for the detection and analysis of
the photon pairs generated by SFWM and the photons generated by
stimulated FWM. The first setup, shown in
Fig.~\ref{Fig:ExperimentalSetup}(c)(i), was used to measure the quantum
correlations from SFWM by coincidence detection.  
A 99:1\% coupler was used to monitor the 1\% output
power exiting the SiNW via a PM. The remaining 99\% was sent to a multi-output
liquid-crystal-on-silicon waveshaper (LCoSWS, Finisar Waveshaper) that
separated idler and signal photons into distinct spatial mode channels. The two
channels were then broadband filtered to remove any residual pump photons
before entering another two PCs inserted before the two superconducting single
photon detectors (SSPDs, Single Quantum) to optimize the detection efficiency
of the two channels. Coincidence measurements were conducted and recorded by a
computer via a time interval analyzer (TIA, SensL). 
The spectral resolution obtained in each channel was 10~GHz, limited by the pixel
bandwidth of the LCoSWS. This led to a $40\times40$ pixel grid for the final JSI. 

The next setup, shown in Fig.~\ref{Fig:ExperimentalSetup}(c)(ii), implemented the singles-based characterization
of the JSI. In addition to the pump pulse, to stimulate FWM the seed laser described in
Fig.~\ref{Fig:ExperimentalSetup}(a)(iii) was also injected into the SiNW at
higher frequency than the pump, corresponding to the measured signal band
in SFWM measurements. The generated average
power in the idler band was approximately 1.8~$\mu$W. 
Instead of performing coincidence measurements, 
we measured the singles count rate recorded by one SSPD in the 
idler detection band,
with the seed laser operating in the signal band.
Both the seed laser frequency and the idler detection band (controlled by the LCoSWS)
were scanned in 10~GHz units in a raster fashion.
Again, the spectral resolution obtained was 10~GHz
with the extracted JSI represented  on 
a $40\times40$ pixel grid.

A 20~dB attenuation was applied in the LCoSWS to limit
the rate of idler photons being detected by the SSPD, thus avoiding saturation.
The final measurement setup, shown in Fig~\ref{Fig:ExperimentalSetup}(c)(iii),
is the OSA measurement of the JSI using stimulated FWM. In this measurement we
kept the scanning seed laser as in the singles-based measurement, but replaced
the LCoSWS, PC, SSPD and TIA with an optical spectrum analyser (OSA, Yenista)
that provided a higher resolution of 2.5~GHz. 
The resulting JSI has four times higher resolution with a $157\times157$ grid.

\subsection*{Theoretical calculations }

To theoretically model the JSA for the different laser pulses, 
the SiNW dispersion relation was approximated as
\begin{equation}
k(\omega) = k(\omega_{p})+\frac{1}{v_p}(\omega-\omega_{p})+
\frac{\beta_{2}(\omega_{p})}{2}
(\omega-\omega_{P})^{2},
\end{equation}
\noindent where $k(\omega_{p}) = 9.63\times10^{6}\ \mathrm{m^{-1}}$, $v_{p} = 7.02\times10^{7}\
\mathrm{m/s}$ and $\beta_{2}(\omega_{p}) = -6.03\times10^{-25}\
\mathrm{s^{2}/m}$. Using these parameters and the geometry
of the waveguide, we used Eq.~\eqref{eq:jsatheory} to calculate the expected 
JSA and JSI for the two laser pulses. The resulting JSI distributions  
are shown in Fig.~\ref{Fig:Results}(a)(i) and
Fig.~\ref{Fig:Results}(b)(i). As expected, for pulses increasing in duration towards quasi-CW, the high SNLB in
Fig.~\ref{Fig:Results}(a)(i) indicates a more entangled state compared with
Fig.~\ref{Fig:Results}(b)(i).

\subsection*{Limitations }

Unlike some SPDC and SFWM schemes that are phase-matched far from the pump, our
SiNW dispersion does not allow for measurement of the whole JSI as the pump frequency
lies in the centre (see Fig.~\ref{Fig:Concept} JSI plot). However this is not a
serious restriction, since this band will also be inaccessible in any
application of such a source.  Our measurement is therefore concerned with an
experimentally accessible portion of the JSI, over a tuning
range of 0.745--1.135~THz (5.94--9.15 nm) from the center frequency of the pump.
We use the SNLB as a measure of the accuracy with which the JSI is
extracted in each case. However, the measured values of $\tilde{K}$ are
affected by the available frequency resolution, as well as the noise
in each class of measurement.
To understand the impact of limited resolution, and thus separate this from the impact of noise in the experimental data, for each of the pump
pulse lengths, we calculated the expected theoretical
values of $\tilde{K}$,  at each of the available frequency resolutions
and a reference value at much finer resolution beyond which $\tilde{K}$ does not change in the 4th decimal place. Note that in our case, the combination of 
accessible frequency range and dispersion strength meant that no difference was found 
between the value of the SNLB, $\tilde{K}$, and the true Schmidt number, $K$, in the 
high resolution calculations. This would not be true in general of course.

The expected impact of limited resolution is shown in Table~\ref{tab:theoryschmidt}.
It is clear that for the narrow bandwidth 270~ps source, the maximum
observable value of $\tilde{K}$ is significantly reduced from its
ideal value.  On the other hand, for the broadband 10~ps source, 
even the coarse $40\times 40$ grid can represent a $\tilde{K}$ exceeding
80\% of the ideal value.
The extent to which the measured values fall below these limits is a measure
of the impact of noise of various types.

\begin{table}[htbp]
\centering

\begin{tabular}{c|cccc}
\begin{tabular}{@{}c@{}}Pump\\ Pulse Duration\end{tabular} & 
       \begin{tabular}{@{}c@{}}40 by 40\\ grid $\tilde{K}$ \end{tabular}&  
       \begin{tabular}{@{}c@{}}157 by 157\\ grid $\tilde{K}$ \end{tabular}  &
       Ideal $\tilde{K}$  & \\
\hline
270 ps  & 38.99 & 83.52   & 95.27\\
10 ps & 8.09 & 8.41     & 9.97\\
\end{tabular}
\caption{ \footnotesize{Theoretically extracted SNLBs for a 40 by 40, 157 by 157 and ideal JSIs, 
for the 270 ps 100 MHz and 10 ps 50 MHz laser pump pulses.}}
\label{tab:theoryschmidt}
\end{table}

\section*{Results}

\begin{figure*}[t]
\centering
\includegraphics[width=15.5cm]{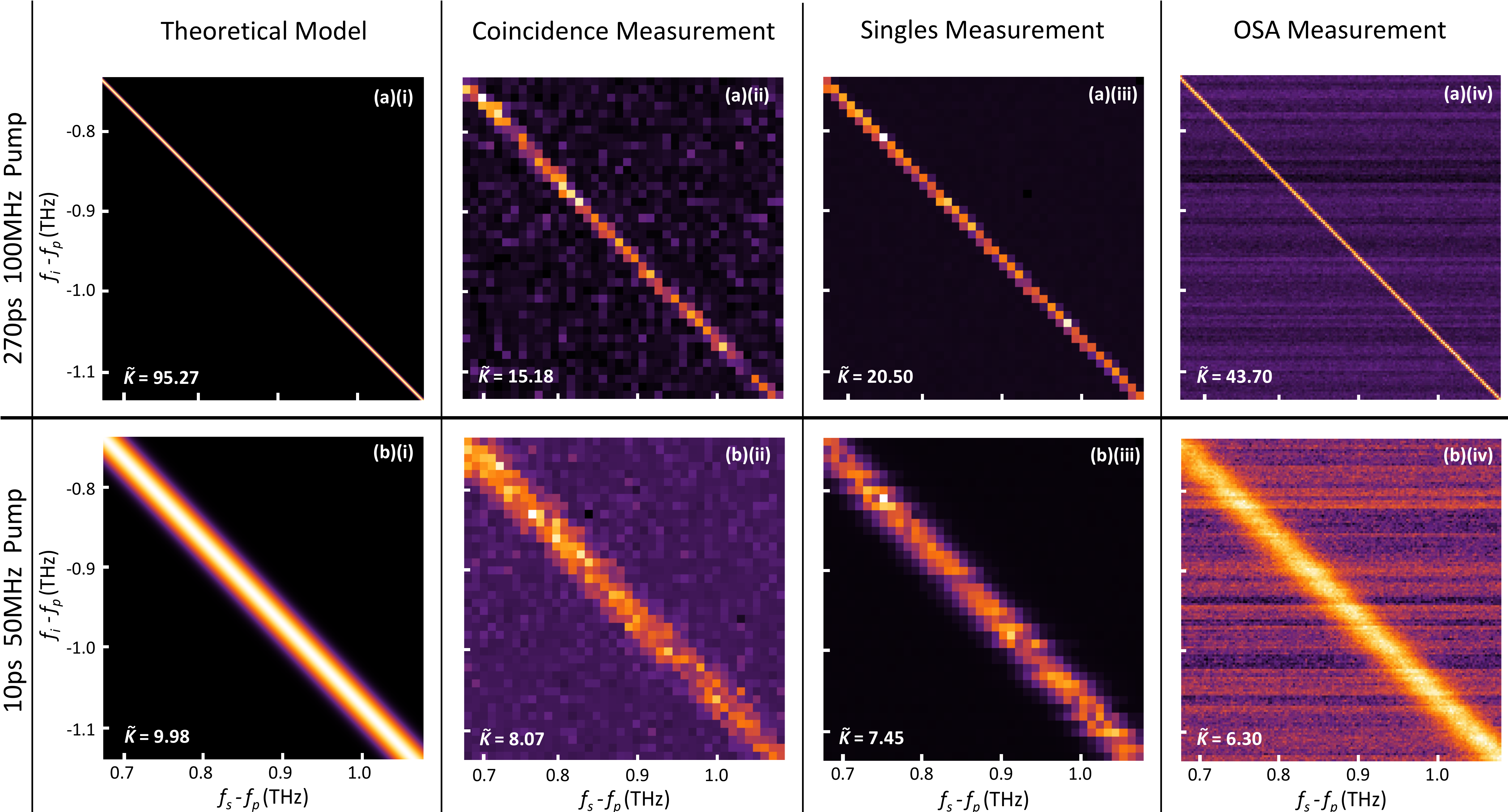}
\caption{\footnotesize{The theoretically-calculated model and results of the six JSI measurements, for (a) the 270~ps and (b) 10~ps pump laser pulses: (i) theoretical ideal model, (ii) photon pair coincidence measurement, (iii) stimulated FWM singles-based measurement and (iv) stimulated FWM OSA measurement.}}
\label{Fig:Results}
\end{figure*}

With the combination of the two laser pulses and the three detection methods, we
measured a total of six partial JSIs. The theoretical and the three
experimental JSI measurements are shown in Fig.~\ref{Fig:Results}(a) and (b)
for the 270~ps and 10~ps pulses respectively, with their associated SNLBs
 $\tilde{K}$ estimated by singular value decomposition.

\end{multicols}

\begin{multicols}{2}

The total time taken to build up the 40 by 40 pixel grid (10~GHz resolution)
coincidence JSI plots shown in Figs.~\ref{Fig:Results}(a)(ii)
and~\ref{Fig:Results}(b)(ii) was approximately 36 hrs and 33 hrs respectively.
During this time we continually adjusted the LCoSWS pass band for each channel
across the whole JSI at a rate of 6 pixels per minute, summing the pixels from
each scan until the largest number of recorded coincidence counts in any one
pixel was 105.  This repeated sampling process was
designed to minimize the effect of slow fluctuations in laser power and
waveguide couplings. As theoretically predicted, the broader spectral profile
$\alpha(\omega)$ of the 10~ps laser source results in a broader anti-diagonal
band, and thus a lower SNLB, for its associated JSI than that associated with
the 270~ps source. As this is a SFWM measurement, the impact of accidental coincidences in JSI plots is large and contributes to a lower SNLB than predicted.


The generated single photon measurements corresponding to the experimental
setup in Fig.~\ref{Fig:ExperimentalSetup}(c)(ii) are plotted in
Fig.~\ref{Fig:Results}(a)(iii) and Fig.~\ref{Fig:Results}(b)(iii). As
stimulated FWM leads to a count rate at a single detector on the order of
$\mathrm{10^{5}~s^{-1}}$, very low relative numbers of background singles are
seen when scanning the LCoSWS band pass filter. The counts in the dark
background region are only limited by dark counts from our detectors, which are
on the order of $\mathrm{100~s^{-1}}$.  

This high signal-to-noise ratio in turn results in a higher SNLB being obtained
when compared with the coincidence measurements, evident in the 270~ps pumped
singles measurement in Fig.~\ref{Fig:Results}(a)(iii). 
However, a slightly lower SNLB was obtained for the 10~ps laser pulse in Fig.~\ref{Fig:Results}(b)(iii)
compared with the corresponding coincidence measurement.  This is caused by the
non-uniform distribution of singles across the anti-diagonal band of the plot
which is a result of small fluctuations in the laser powers, detector
efficiency, and polarization from scan to scan. Additionally, 
the 10 ps pumped coincidence value for the $40 \times40$ grid ($\tilde{K}=8.09$)  
provided the closest agreement to the expected value ($\tilde{K} = 8.07$) 
when compared with the singles-based measurement. Due to the
high rate of stimulated FWM idler photon generation, the integration time for
both pump measurements  was limited only by the scanning speed of the seed
laser and the LCoSWS, as well as the speed of the electronic acquisition. Thus,
the fastest possible integration time for both measurements was only 1.5 hours.
Still, moving to this singles-based measurement results in a significant decrease
in the required integration time when compared to the coincidence measurement,
while providing comparable SNLBs.


The classical OSA measurements shown in Fig.~\ref{Fig:Results}(a)(iv) and
Fig.~\ref{Fig:Results}(b)(iv) were completed within 2 hours for each laser
pulse width but with $16$ times higher resolution, at the maximum resolution of 2.5~GHz.The horizontal streaks visible in both JSI plots are a result of the constant change in the noise floor
of the OSA with every trace measurement. In theory, the streaks can be
eliminated by averaging multiple traces for a fixed seed probe, but not without
increasing the total integration time of each JSI measurement. In principle,
using this method we are able to measure the complete JSI profile of the SiNW,
as the OSA is not saturated by the input pump at the powers used here.

\section*{Conclusion and Outlook} We have presented measurements
comparing JSIs from a $\chi^{(3)}$ nonlinear device, in our case a SiNW, via three different experimental methods that can be used to characterize the
entanglement between generated photon pairs. This is achieved by employing both
quantum correlation measurements and classical stimulated measurements, which makes use of the relationship
between SFWM and stimulated FWM.  For the stimulated FWM processes, we have shown two
techniques, one that uses no further components than quantum correlations,
other than a CW probe laser, and the other using a high resolution OSA. By
successfully measuring the JSI for two different laser pulses, we observed a
direct change in the spectral entanglement of the generated photon states,
proving the versatility of our characterization schemes. For the JSI
measurements, we saw by switching from the LCoSWS to an OSA, we were able to
increase the resolution from 10~GHz to 2.5~GHz, however this also resulted in
horizontal streaks in the JSI, a problem attributed to the change in the noise
floor of the OSA. In the future this could be overcome by using a lower noise
OSA, using an OSA with an output for a single photon detector, or limiting
measurements to nonlinear devices with a higher FWM conversion efficiency.  By
comparing the SNLBs calculated via SVD of our experimental
measurements with our ideal theoretical model, we conclude that the OSA
provided the most accurate spectral entanglement measurement (although half of
the predicted SNLB) for the long pump pulse. However, the measured
spectral entanglement for the short pump pulse via the OSA provided us with the
largest deviation from the ideal model, with the results obtained via
coincidence measurement being in closest agreement with the theory.
Nonetheless, the OSA measurement will consistently provide the fastest and
highest resolution for future measurements of JSIs. Overall, the long pump
pulse spectral entanglement measurement provided us with the biggest
discrepancy when compared to the ideal model, caused by the discretized JSI
measurements having a limited resolution at the same scale as the pump spectral
profile. In the future these methods could be applied to other integrated pair
generation devices including ring resonators~\cite{Clemmen2009,Engin2013} and
slow-light photonic crystals~\cite{Xiong2011, Xiong2012, Clark2013Chad,
Matsuda2013}.  The methods presented here are of substantial importance for
future characterization of spectrally complex two photon states, particularly
for nonlinear devices that require fast and reliable measurements, and when
large numbers of devices must be characterized for use in future quantum
technologies.

\section*{Acknowledgments}

This work was supported by the Centre of Excellence (CUDOS, project number CE110001018), Laureate Fellowship (FL120100029), Future Fellowship (FT110100853), Discovery Project (DP130100086) and Discovery Early Career Researcher Award programs (DE130101148 and DE120100226) of the Australian Research Council (ARC).  L.G.H. acknowledges support from a Macquarie University Research Fellowship.





\bibliographystyle{unsrt}

\bibliography{My_Bib}
\end{multicols}
\end{document}